\documentclass[12pt]{iopart}

\usepackage{iopams}

\usepackage{graphicx}

\def\Journal#1#2#3#4#5#6{#1, ``#2'', {\em #3} {\bf #4}, #5 (#6).}

\def\JournalPrep#1#2#3{#1, ``#2'', #3.}
\newcommand\bm[1]{\mbox{\boldmath$#1$}}

\def\JDG{J. Diff. Geom.}
\def\CQG{Class. Quantum Grav.}

\def\JMP{J. Math. Phys.}

\def\CMP{Comm. Math. Phys.}

\def\ANYAS{Ann. N. Y. Acad. Sci.}

\def\APP{Acta Phys. Pol.}

\def\T{\cal T}

\begin{document}

\newtheorem{defi}{Definition}
\newtheorem{lema}{Lemma}
\newtheorem{proof}{Proof}
\newtheorem{thr}{Theorem}
\newtheorem{corollary}{Corollary}
\newtheorem{proposition}{Proposition}

\def\L{\mathcal{L}}
\def\Db{{\mathfrak D}_{b}}
\def\D{\mathfrak D}\def\Sb{S_{b}}
\def\S{S}
\def\U{U}
\def\Sm{S_{2}}
\def\Sf{S_{f}}

\def\U{U}
\def\ycoor{\hat{y}}
\def\y{y}
\def\Y{Y}
\def\P{P}
\def\Q{Q}
\def\Pcal{\mathcal{P}}
\def\Qcal{\mathcal{Q}}
\def\f{f}
\def\F{F}
\def\Ysol{Y_1}
\def\B{B_{Y,\epsilon}}
\def\Bn{B_{Y_{n},\epsilon_{n}}}

\title{A counter-example to a recent version of the Penrose conjecture}
\author{Alberto Carrasco and Marc Mars}
\address{Dept. F\'{\i}sica Fundamental, 
Universidad de Salamanca, Plaza de la Merced s/n, 37008 Salamanca, Spain}
\eads{\mailto{acf@usal.es} and  \mailto{marc@usal.es}}
\begin{abstract} 
By considering suitable axially symmetric slices on the Kruskal 
spacetime, we construct a counterexample to a recent version of
the Penrose inequality in terms of so-called generalized apparent horizons. 
\end{abstract} 


\vspace{5mm} 

\section{Introduction}

In a serious attempt \cite{BrayKhuri2009} to prove the Penrose inequality in the general case,
Bray and Khuri were led to conjecture a new version of the Penrose inequality
in terms of so-called generalized apparent horizons.  
In fact, they proved
that if a certain system of PDEs admit solutions with the right 
boundary behaviour, then such a Penrose inequality follows. In this paper
we show that this inequality cannot be true in general by finding 
slices of the Kruskal spacetime (i.e. the maximally extended Schwarzschild spacetime) 
for which the outermost generalized
apparent horizon has area strictly larger than $16 \pi M^2$, where $M$
is the ADM mass of the spacetime. We start with a brief discussion 
on the Penrose inequality, with the aim of putting the Bray and Khuri 
proposal into context (see \cite{Mars2009} for further details) and then show
that there exist slices of Kruskal for which this inequality is violated. 
For the systems of PDEs proposed in \cite{BrayKhuri2009}, this means that
a general existence theory cannot be expected with boundary conditions compatible with 
generalized apparent horizons. However, simpler boundary conditions (e.g.
compatible with
future and past apparent horizons) are not ruled out. This may in fact
simplify the analysis of these equations.


Penrose \cite{Penrose1973} noticed that the total mass
of a spacetime containing black holes that settle down to a stationary
state must satisfy the inequality 
\begin{equation}\label{penrose0}
M_{ADM}\geq \sqrt{\frac{|\mathcal{H}|}{16\pi}},
\end{equation}
where $|{\cal H}|$ is the area of the event horizon
at one instant of time. Moreover, assuming 
the matter contents to satisfy the dominant
energy condition and combining (\ref{penrose0}) with 
cosmic censorship, Penrose observed that new inequalities 
similar to (\ref{penrose0}) follow, where the right-hand side
is replaced by the area of certain surfaces which can be defined
independently of the future evolution of the spacetime (in contrast
to the event horizon). This type of inequalities are collectively
termed {\it Penrose inequalities}. Their main interest 
is two-fold. Firstly, they provide
strengthenings of the positive mass theorem. 
Secondly, since cosmic censorship
is the basic physical ingredient supporting their
validity, a direct proof
of the inequality would give rather strong indirect support 
for the cosmic censorship conjecture.

There are several versions of the Penrose inequality. Typically,
one considers closed (i.e. compact and without boundary) surfaces $S$ embedded
in a spacelike and asymptotically flat hypersurface $\Sigma$, which
are {\it bounding}, i.e. such that
$S$ divides $\Sigma$ into two open regions. The region containing the 
asymptotically flat end is called ``exterior'', while it complementary 
is the ``interior'' of $S$. 
Given two bounding surfaces $S_1$ and $S_2$, we say that $S_1$ {\it
encloses} $S_2$ provided the exterior of $S_2$ contains the exterior of $S_1$.
We denote by $\theta^{+}$ and $\theta^{-}$ the null expansions along the
outer and inner future null directions respectively. A surface is called
{\bf weakly outer trapped} iff $\theta^{+} \leq 0$
and
{\bf marginally outer trapped surfaces (MOTS)} if $\theta^{+}=0$. In terms
of the initial data set $(\Sigma,g_{ij},K_{ij})$,
we have
$\theta^{\pm} = p \pm q$, where $p$ is the mean curvature of $S \subset (\Sigma,g_{ij})$
with respect to the outer unit 
normal $\vec{m}$ and $q$ is the trace of the pull-back of the second fundamental 
form $K_{ij}$ onto $S$.
The union of the
interiors of all weakly outer trapped surfaces defines the so-called
outer trapped set $\T^+$ in $\Sigma$. The boundary of this set is a smooth
MOTS \cite{AnderssonMetzger2007}.

The standard version of the Penrose inequality reads
\begin{equation}\label{penrose1}
M \geq \sqrt{\frac{A_{\mbox{min}}(\partial \T^{+})}{16\pi}},
\end{equation}
where $A_{\mbox{min}}(\partial \T^{+})$ is the minimal area necessary to enclose
$\partial \T^{+}$. The need of taking this {\bf minimal area enclosure}
comes from the fact that, under cosmic censorship, we know that
the event horizon encloses $\partial \T^{+}$. However, the former could still have less
area than $\partial \T^{+}$ and, since its location is undetermined,
the minimum of area of all enclosing surfaces should be taken.
Inequality (\ref{penrose1}) also has a rigidity statement, namely that 
equality implies that $(\Sigma,g_{ij},K_{ij})$ is a slice
of the Kruskal spacetime.

By reversing the time orientation, 
the same argument yields
(\ref{penrose1}) with $\partial \T^{+}$
replaced by $\partial \T^{-}$, where $\T^{-}$ is the union of the interiors
of all bounding surfaces satisfying $\theta^{-} \geq 0$. In  general,
neither $\partial \T^{+}$ encloses $\partial \T^{-}$ nor
vice versa. In the time-symmetric case $K_{ij}=0$, the inequality simplifies because $\T^{+} = \T^{-}$ and its boundary is
the outermost minimal surface (i.e. a minimal surface enclosing
any other bounding minimal surface 
in $\Sigma$), and hence its own minimal area enclosure.
The inequality in this case is called
{\it Riemannian Penrose inequality} and it has been proven for one
black hole in \cite{HuiskenIlmanen2001} and in full generality in \cite{Bray2001}
using a different method. 
In the non-time symmetric case, (\ref{penrose1}) is not invariant
under time reversals. Moreover, the minimal area
enclosure of a given surface $S$ can be a rather complicated object 
typically consisting of portions of $S$ together with portions of minimal surfaces
(i.e. surfaces with $p=0$) outside of $S$. This complicates the problem 
substantially. This has led several authors to 
propose simpler looking versions of the inequality, even if they
are not directly
supported by cosmic censorship. Two such extensions are 
\begin{eqnarray}
M \geq \sqrt{\frac{A_{\mbox{min}}( \partial (\T^{+} \cup \T^{-} ))}{16\pi}},
\hspace{2cm}  M \geq \sqrt{\frac{|\partial (\T^{+} \cup \T^{-} )|}{16\pi}},
\label{penrose2}
\end{eqnarray}
where $|S|$ denotes the area of $S$ (cf. \cite{Karkowski-Malec2005}).
These inequalities are immediately stronger
than (\ref{penrose1}) 
and have the advantage of being invariant under time
reversals. The second avoids even the use of minimal area enclosures. Neither
version  is supported by
cosmic censorship and at present there is little evidence for their validity. 
However, both reduce to the standard version in the Riemannian case and
both hold in spherical symmetry. No counterexamples 
are known either. It would be interesting to have either stronger 
support for them, or else to find a
counterexample.

Recently, Bray and Khuri proposed a new method to
approach the general (i.e. non time-symmetric) Penrose inequality. The
basic idea was to modify the Jang equation \cite{Jang1978}, \cite{SchoenYau1981}
so that
the product manifold $\Sigma \times \mathbb{R}$ used to construct the graphs which define
the Jang equation is endowed with a
warped type metric of the form $-\varphi^2 dt^2 + g$ instead of the product metric.
The aim was to reduce the general Penrose inequality to 
the Riemannian Penrose inequality on the graph manifold. A discussion on the 
type of divergences that could possibly occur for the generalized 
Jang equation led the authors
to consider a new type of bounding surfaces called
{\bf generalized trapped surfaces} and {\bf generalized apparent horizons}, which
are 
defined, respectively, by $p \leq |q|$ and
$p = |q|$.  This type of surfaces have very interesting properties. 
The most notable one is that, on any asymptotically flat initial data set
containing at least one generalized trapped surface, there is always a
unique $C^{2,\alpha}$ outermost generalized apparent horizon $S_{out}$ \cite{Eichmair2008}. 
Moreover, this surface
has smaller area than any other surface enclosing it \cite{Eichmair2008}. 
Bray and Khuri's version of the Penrose inequality reads
\begin{eqnarray}\label{penroseBK}
M\geq\sqrt{\frac{|S_{out}|}{16\pi}}.
\end{eqnarray}
This inequality has several
remarkable properties that makes it very appealing \cite{BrayKhuri2009}.
First of all, the definition
of generalized apparent horizon, and hence the corresponding Penrose inequality,
is insensitive to time reversals. Moreover, there is no need of taking
the minimal area enclosure of $S_{out}$, as this surface has less area than any of
its enclosures. Since MOTS are automatically generalized
trapped surfaces, $S_{out}$ encloses the outermost MOTS $\partial \T^{+}$. Thus, (\ref{penroseBK})
is stronger than (\ref{penrose1}) and its proof would also 
establish the standard version of the Penrose inequality. Moreover, Khuri has proven
\cite{Khuri2009-2}
that no generalized trapped surfaces exist in Minkowski, which is a necessary condition
for the validity of (\ref{penroseBK}). Another interesting property of this version,
and one of its motivations discussed in \cite{BrayKhuri2009}, is that the
 equality case in (\ref{penroseBK}) 
covers a larger number of slices of Kruskal than the equality case in (\ref{penrose1}). 
Recall that the rigidity statement of any
version of the Penrose inequality asserts that equality implies that $(\Sigma,g_{ij},K_{ij})$ is
a hypersurface of Kruskal. However, {\it which} slices of Kruskal satisfy the equality case may depend
on the version under consideration. The more slices have this property, the more accurate the version can
be considered. For any slice $\Sigma$ of Kruskal we can define its {\it exterior region} $\Sigma^{+}$ as
the intersection of $\Sigma$ with the domain of outer communications.  Bray and Khuri noticed that
whenever $\partial \Sigma^{+}$ intersects both the black hole and the white hole event horizons,
then the standard version (\ref{penrose1}) gives, in fact, a strict inequality. Although (\ref{penroseBK})
does not give equality for all slices of Kruskal, it does so in all cases where the boundary
of $\Sigma^{+}$ is a $C^{2,\alpha}$ surface (provided this boundary -- which is a generalized apparent horizon in this
case -- is, in fact, the outermost such horizon). It follows that version (\ref{penroseBK}) contains more
cases of equality than (\ref{penrose1}) and is therefore more accurate. It should be stressed that the second inequality in 
(\ref{penrose2}) gives equality for
{\it all} slices of Kruskal, so in this sense it would be optimal.

Despite its appealing properties, (\ref{penroseBK}) is {\it not} directly supported
by cosmic censorship. The reason is that the outermost generalized apparent horizon need not always
lie inside the event horizon. A simple example \cite{Mars2009}
is given by a slice $\Sigma$ of Kruskal such that
$\partial \T^{+}$ (which corresponds to the intersection of $\Sigma$ with the black hole event
horizon) and $\partial \T^{-}$ (the intersection $\Sigma$ with the white hole horizon)
meet transversally. Since both surfaces are generalized trapped surfaces,
Eichmair's theorem \cite{Eichmair2008}
implies that there must exist a unique $C^{2,\alpha}$ outermost
generalized apparent horizon enclosing
both. This surface must therefore penetrate into the exterior region somewhere, as claimed.
It becomes natural to study the outermost generalized apparent horizon in slices of this type
in order to check whether (\ref{penroseBK}) holds or not. The result is that there are examples
for which (\ref{penroseBK}) turns out to be violated. More precisely, in this paper
we prove the following statement.

\begin{thr}\label{theorem}
In the Kruskal spacetime with mass $M>0$, there exist asymptotically flat, spacelike
hypersurfaces with an
outermost generalized apparent horizon $S_{out}$ satisfying $|S_{out}|>16\pi M^{2}$.
\end{thr}

\section{Construction of the counterexample.}

Let us  
consider the Kruskal spacetime of mass $M>0$ with metric
\[
ds^2= \frac{32M^{3}}{r}e^{-r/2M}d\hat{u}d\hat{v}+ r^2 \left( d\theta^2 + \sin{\theta}^2 d\phi^2 \right),
\]
where $r(\hat{u}\hat{v})$ solves the implicit equation
$\hat{u}\hat{v}= e^{r/2M}(r-2M)/(2M)$. In this metric 
$\partial_{\hat{v}}$ is future-directed and $\partial_{\hat{u}}$ is past-directed. The 
region $\{\hat{u}>0,\hat{v}>0\}$ defines the  domain
of outer communications and
$\{\hat{u}=0\}$, $\{\hat{v}=0\}$ define, respectively, the black hole and white hole
event horizons.
Consider the one-parameter family of axially-symmetric embedded hypersurfaces
$\Sigma_{\epsilon} = \mathbb{R} \times S^2$, with intrinsic coordinates $\ycoor \in \mathbb{R}$, $x \in [-1,1]$,
$\phi \in [0,2 \pi]$, defined by the embedding
\begin{eqnarray*}
\Sigma_{\epsilon}\equiv \left\{ \hat{u}=\ycoor- \epsilon  x, \hat{v}=\ycoor+ \epsilon x,
\cos \theta = x , \phi = \phi\right\}.
\end{eqnarray*}
It is easy to check that these hypesurfaces are well-defined, smooth and asymptotically flat for all 
$|\epsilon| < 1$. Morever, it is straightforward to show that $\Sigma_{\epsilon}$ is spacelike everywhere
for $|\epsilon|$ small enough.  The discrete isometry of the Kruskal
spacetime defined by $\left\{\hat{u},\hat{v}\right\}\rightarrow 
\left\{\hat{v},\hat{u}\right\}$ implies 
that under reflection with respect to the equatorial plane, i.e. $(\ycoor,x,\phi) \rightarrow (\ycoor,-x,\phi)$,
the induced metric of $\Sigma_{\epsilon}$ remains invariant, while the second fundamental
form of $\Sigma_{\epsilon}$ changes sign.
The exterior region $\Sigma^{+}_{\epsilon} $ of $\Sigma_{\epsilon}$ 
is given by $\{\ycoor - | \epsilon x |>0\}$. For
$\epsilon  \neq 0$, $\partial \Sigma^{+}_{\epsilon}$ is composed by a portion of the
black hole event horizon and a portion of the white hole event horizon. Moreover, $\partial \T^{+}$
is given by $\{\ycoor - \epsilon x=0\}$, while $\partial \T^{-}$
is $\{\ycoor + \epsilon x=  0 \}$ so that these surfaces intersect transversally on the circumference
$\{\ycoor=0, x= 0\}$ provided $\epsilon \neq 0$. By Eichmair's theorem, there exists
a $C^{2,\alpha}$ outermost generalized apparent horizon containing both $\partial \T^{+}$ and
$\partial \T^{-}$. Uniqueness implies that this surface must be axially symmetric and have equatorial symmetry. 
In order to locate it, we proceed in two steps. First we will show
that an axial and equatorially symmetric generalized apparent horizon of spherical topology and lying 
in a sufficiently small neighbourhood of $\{\ycoor=0\}$  exists, provided 
$\epsilon$
is small enough. We will also
determine its embedding function
to first order in $\epsilon$. In the second step we prove that this surface 
is either the outermost generalized apparent horizon, or else, it has smaller 
area than any other
generalized apparent horizon in $\Sigma_{\epsilon}$ enclosing it.

Thus, we consider surfaces of spherical topology
defined by embedding functions
$\{ \ycoor = \y (x, \epsilon), x=x, \phi = \phi \}$ and satisfying $\y(-x,\epsilon)=\y(x,\epsilon)$.
Since the outermost generalized apparent horizon is known to be $C^{2,\alpha}$ it is natural to 
consider the spaces of functions
$U^{m,\alpha} \equiv 
\left\{ \y\in C^{m,\alpha}(S^2):\partial_{\phi}\y=0,\y(-x)= \y(x)\right\} $,
i.e. the spaces of $m$-times differentiable functions on the unit sphere, with H\"older continuous 
$m$-th derivatives with exponent $\alpha \in (0,1)$ 
and invariant under the axial Killing vector
on $S^2$ and under reflection about the equatorial plane. Each space $U^{m,\alpha}$ is a closed
subset of the Banach space $C^{m,\alpha}(S^2)$ and hence a Banach space itself. 
Let $I \subset \mathbb{R}$ be the closed interval where $\epsilon$ takes values.
For each function 
$\y \in U^{2,\alpha}$
the expression $p - |q|$ defines a non-linear map 
$\f:U^{2,\alpha}\times I \rightarrow U^{0,\alpha}$. Thus, we are looking for the outermost of the 
solutions $\y\in U^{2,\alpha}$ of the equation $\f=0$. We know that when $\epsilon=0$,
the hypersurface $\Sigma_{\epsilon}$ 
is totally geodesic, which implies $q=0$ for any surface on it. 
Consequently, all
generalized apparent horizons on $\Sigma_{\epsilon=0}$ satisfy $p=0$ and are, in fact, 
minimal surfaces. The only closed minimal surface in $\Sigma_{\epsilon =0}$ is 
the bifurcation surface $\left\{ \hat{u}=0, \hat{v}=0 \right\}$. Thus, the equation 
$\f(\y,\epsilon)=0$ has $\y=0$ as the unique solution when $\epsilon=0$.
It becomes natural to use the implicit function theorem for Banach spaces to show that 
there exists a unique solution $\y\in U^{2,\alpha}$ of $\f=0$ in a neighbourhood of $\y=0$ for $\epsilon$ small enough.
The implicit function theorem requires the operator $\f$ to have a continuous Fr\'echet derivative and the 
partial derivative $\left.D_{\y}\f\right|_{(\y=0,\epsilon=0)}$ to be an isomorphism. 
The appearance of an absolute value in $|q|$ makes the Fr\'echet derivative of $\f$ potentially 
discontinuous \cite{Khuri2}. 
However, the problem can be solved considering a suitable modification of $\f$. 
Since the details are somewhat technical, we postpone the proof of this fact to an Appendix, 
where we establish the followig proposition.

\begin{proposition}\label{proposition}
There exists a neighborhood $\tilde{I}\subset I$ of $\epsilon=0$ 
such that $\f(\y,\epsilon)=0$ admits a solution
$\y(x,\epsilon) \in C^{2,\alpha}(S^2)$  for all $\epsilon \in \tilde{I}$.
Moreover, $\y(x,\epsilon)$ is $C^1$ in $\epsilon$
and satisfies $\y(x,\epsilon=0)=0$.
\end{proposition}

Let us denote by $S_{\epsilon}$ the surface defined by this solution.
The proposition above implies that we can expand $\y(x, \epsilon)=
Y_{1}(x)\epsilon + o(\epsilon)$.
By linearizing the PDE $\f(y,\epsilon)=0$ in $\epsilon$, it turns out
that $Y_{1}$ satisfies the linear equation $L(Y_{1}(x))= 3 |x|$, where 
$L(z(x))\equiv -(1-x^2)\ddot{z}+2x\dot{z}+z$. The right hand side 
of this equation corresponds (except for a positive multiplicative constant) to the linearization of $|q|$.
By decomposing into Legendre polynomials $P_n(x)$, it is easy to show that the unique solution of this
equation is
\begin{eqnarray*}
\hspace{-1cm} Y_1(x)  =   \frac32 + \sum_{n=1}^{\infty}a_{2n}P_{2n}(x), \quad
a_{2n}   =  \frac{3(4n+1) (-1)^{n+1}}{\left[ 2n(2n+1)+1 \right]2^{2n}}\frac{(2n-2)!}{(n-1)!(n+1)!},
\end{eqnarray*}
where convergence is in $L^2$. This expression allows us to compute the area
of $S_{\epsilon}$ at once. After a straightforward calculation we find
\begin{equation*}
|S_{\epsilon}| = 16 \pi M^2 +  \frac{8  \pi M^2 \epsilon^2}{e} \left ( 
5 + 4 \sum_{n=1}^{\infty}  \frac{2n(2n+1)+1}{4n+1}a_{2n}^2 \right) + O (\epsilon^3).
\end{equation*}
Since the second summand is strictly positive, it follows that $|S_{\epsilon}| > 16 \pi M^2$. 
If we could show that $S_{\epsilon}$ is the outermost generalized apparent horizon, we would
have a counterexample of (\ref{penroseBK}). Before turning into this point, however,
let us give an alternative argument to show that the area increases. This will shed some light into
the underlying reason why the area of $S_{\epsilon}$ is larger than $16 \pi M^2$.

To that aim, let us now use coordinates $\{\hat{u},x,\phi\}$ in $\Sigma_{\epsilon}$. 
Then, the embedding of $\Sigma_{\epsilon}$ becomes 
$\Sigma_{\epsilon}\equiv\left\{ \hat{u},\hat{v}=\hat{u}+2 \epsilon x,x,\phi \right\}$, and the
corresponding embedding 
in $\Sigma_{\epsilon}$ for the surfaces $S_{\epsilon}$
is $S_{\epsilon}=\left\{ \hat{u} = u(x,\epsilon),x,\phi \right\}$. Again,
$u$ admits an expansion $u = U_1(x)  \epsilon + o(\epsilon)$. The
relationship between $U_1$ and $Y_1$ is simply $Y_{1}=U_{1}+x$. It follows
that $U_1$ satisfies $L(U_{1}(x))= 3(|x|- x)$.
Similarly, if we take $\left\{ \hat{v},x,\phi \right\}$ as coordinates for $\Sigma_{\epsilon}$,
then  the embedding of $S_{\epsilon}$ reads $\hat{v} = V_1 (x) \epsilon + o (\epsilon)$, with
$V_1$ satisfying the equation
$L(V_1(x))= 3 (|x|+ x)$.
Thus,  $L(U_{1}(x))\geq 0$ and $L(V_1(x)) \geq 0 $ and neither of them is identically zero.
Since $L$ is an elliptic operator with positive zero order term,
we can use the maximum principle 
to conclude that $U_1(x)>0$ and $V_{1}(x)>0$ everywhere. Geometrically, this means 
that $S_{\epsilon}$ lies fully in $\Sigma^{+}_{\epsilon}$ 
for $\epsilon$ small enough. In fact, the maximum principle applied to $L(Y_1) = 3 |x|$
also implies $Y_1 > 0$. This will be used below.

We can now view $S_{\epsilon}$ as a first order spacetime variation of
the bifurcation surface. The variation vector $\partial_{\epsilon}$ is defined 
as the tangent vector to the curve generated when a  point with fixed coordinates $\{x,\phi\}$ in $S_{\epsilon}$ moves
as $\epsilon$ varies. By the argument above, this vector
is spacelike everywhere on the unperturbed
surface $S_{\epsilon=0}$. If we do a Taylor expansion of $|S_{\epsilon}|$ around $\epsilon=0$, 
we see that the zero order term is $|S_{\epsilon=0}| = 16 \pi M^2$, as this is the
area of the bifurcation surface. 
The bifurcation surface is totally geodesic so that, in particular, its 
mean curvature vector vanishes.
Consequently, the linear term in the expansion is identically zero as a consequence
of the first variation of area 
\begin{eqnarray}\label{firstvariationarea}
\frac{d |S_{\epsilon} |}{d \epsilon} = \int_{\S_{\epsilon}} ( \vec{H}_{S_{\epsilon}}, \partial_{\epsilon} )
\bm{\eta_{S_{\epsilon}}},
\end{eqnarray}
where $\vec{H}_{S_{\epsilon}}$ is the spacetime mean curvature vector of $S_{\epsilon}$ and $( \, , \, )$ denotes
scalar product with the spacetime metric.  For the second order term in the expansion, we take the derivative of (\ref{firstvariationarea}) with respect to $\epsilon$ and
evaluate at $\epsilon=0$. A simple computation
gives
\begin{eqnarray*}
\left. \frac{d^2 |S_{\epsilon}|}{d\epsilon^2} \right|_{\epsilon=0}= \frac{16\pi M^2}{e}\int _{-1}^1 \left[ 
\frac{}{} U_1(x)L(V_1(x))+V_1(x)L(U_1(x))\right] dx.
\end{eqnarray*} 
Since $U_1$ and $V_1$ are strictly positive and
$L(U_{1}(x))$, $L(V_1(x))$ are non-negative and not identically zero, it follows
$\left. \frac{d^2 |S_{\epsilon}|}{d\epsilon^2} \right|_{\epsilon=0}>0$ and hence that the area of $S_{\epsilon}$
is larger than $16 \pi M^2$ for small $\epsilon$. The fact that the area
increases is therefore a consequence of the fact that the second order variation of area
turns out to be strictly positive along the direction joining the bifurcation surface with $S_{\epsilon}$,
and, in turn, this is tied to the fact that $L(U_1)$ and $L(V_1)$ have a sign.
The right hand sides of these operators are (except for a constant) the linearization of $|q| \pm q$ 
and these objects are obviously non-negative
in all cases. We conclude, therefore, that the fact that the area of $S_{\epsilon}$ is larger than $16\pi M^2$
is closely related to the defining equation $p = |q|$.
It follows that the increase of area is a robust property
which does not depend strongly on the choice of hypersurfaces $\Sigma_{\epsilon}$ that we have made.
In fact, had we chosen hypersurfaces
$\Sigma_{\epsilon}\equiv \left\{ u=y- \epsilon \beta(x) , v=y+ \epsilon \beta(x)
, \cos \theta = x, \phi = \phi \right\}$,
the corresponding equations would have been
$L(U_{1}(x))=|L(\beta(x))| - L(\beta(x)) $ and
$L(V_{1}(x))=|L(\beta(x))| + L(\beta(x))$. The same conclusions would follow
provided the right hand sides are not identically zero.

Having shown that $|S_{\epsilon}| > 16 \pi M^2$ for $\epsilon \neq 0$
small enough, the next step is to analyze whether $S_{\epsilon}$ is the
outermost generalized apparent horizon or not. In fact, in order to have a counterexample 
of (\ref{penroseBK}) we only need
to make sure that no generalized apparent horizon with less area than
 $S_{\epsilon}$
and enclosing $S_{\epsilon}$ exists in $\Sigma_{\epsilon}$. We will argue by contradiction. 
Let $\hat{S}_{\epsilon}$ be a generalized
apparent horizon enclosing $S_{\epsilon}$ and
with $|\hat{S}_{\epsilon} | < |S_{\epsilon}|$. Then, since $S_{\epsilon}$ is not area outer minimizing,
its minimal area enclosure $S'_{\epsilon}$ does not coincide with $S_{\epsilon}$. Now, two
possibilities arise: (i) either $S'_{\epsilon}$ lies completely outside $S_{\epsilon}$, or (ii)
it coincides with $S_{\epsilon}$ on a closed subset $K_{\epsilon}$, while the complement 
$S'_{\epsilon} \setminus K_{\epsilon}$ (which is non-empty) has vanishing mean
curvature $p$ everywhere. To exclude case (i), consider the foliation of $\Sigma_{\epsilon}$ defined
by the surfaces $\{\ycoor=y_0,x,\phi\}$, where $y_0$ is a constant. A direct computation shows that 
the mean curvature $p_{y_0}$ of these surfaces
with respect to the outer normal is positive for all $y_0 >0$.
We noted above that $Y_1(x) >0$
everywhere. Thus, for small enough $\epsilon$, the function $y(x,
\epsilon)$  is also strictly
positive. Since $S'_{\epsilon}$ lies fully outside $S_{\epsilon}$, the coordinate function $\ycoor$ restricted to
$S'_{\epsilon}$ achieves a positive maximum $y_{\epsilon}$ somewhere. At this point, the two
surfaces $S'_{\epsilon}$ and $\{\ycoor = y_{\epsilon}\}$ meet tangentially, with $S'_{\epsilon}$ lying fully inside
$\{ \ycoor = y_{\epsilon} \}$. This is a contradiction to the maximum principle for minimal
surfaces. It only remains to deal with case (ii). The same argument above shows that the coordinate function
$\ycoor$ restricted to $S'_{\epsilon} \setminus K_{\epsilon}$ cannot reach a local maximum. It follows that 
the range of variation of $\ycoor$ restricted to $S'_{\epsilon}$ is contained in the range of variation of
$\ycoor$ restricted to $S_{\epsilon}$. Since $\max_{S_{\epsilon}} \ycoor -
\min_{S_{\epsilon}} \ycoor = O (\epsilon)$, 
it follows that we can regard $S'_{\epsilon}$ as an outward variation of $S_{\epsilon}$ of order $\epsilon$ when
$\epsilon$ is taken small enough. The corresponding 
variation vector field $\vec{\xi}$ can be taken orthogonal 
to $S_{\epsilon}$ without loss of generality, i.e.  
$\vec{\xi}=\xi \vec{m}$, where $\vec{m}$ is the outward unit normal to $S_{\epsilon}$.
The function $\xi$ vanishes on $K_{\epsilon}$ and is positive in its complement
$U_{\epsilon} \equiv S_{\epsilon} \setminus K_{\epsilon}$. Expanding to second
order and using the first and second variation of area
(see e.g. \cite{Chavel}) gives 
\begin{eqnarray*}
\hspace{-25mm} 
|S'_{\epsilon} |=  |S_{\epsilon}|  +   
\epsilon \int_{U_{\epsilon}}   p_{S_{\epsilon}}\xi 
\bm{\eta_{S_{\epsilon}}} + \\
\hspace{-1cm}  +  \frac{\epsilon^2}{2} 
\int_{U_{\epsilon}} 
\left ( |\nabla_{S_{\epsilon}}\xi|^2 +
\frac{\xi^2}{2}
\left( R^{S_{\epsilon}}-R^{\Sigma_{\epsilon}}-|A_{S_{\epsilon}}|^{2} + p_{S_{\epsilon}}^2
\right )  + p_{S_{\epsilon}} \frac{d\xi}{d\epsilon}  \right )  
\bm{\eta_{S_{\epsilon}}} + O (\epsilon^3),
\end{eqnarray*}
where $\nabla_{S_{\epsilon}}$, $R^{S_{\epsilon}}$ and $A_{S_{\epsilon}}$ are, respectively,
the gradient, scalar curvature and second fundamental form of $S^{\epsilon}$, and $R^{\Sigma_{\epsilon}}$
is the scalar curvature of $\Sigma_{\epsilon}$. Now, the
mean curvature $p_{S_{\epsilon}}$ of $S_{\epsilon}$ reads $p_{S_{\epsilon}} =\frac{3 \epsilon}{M \sqrt{e}}  |x| + o (\epsilon)$
and both $R^{\Sigma_{\epsilon}} $ and $A_{S_{\epsilon}}$ are of order $\epsilon$ (because $\Sigma_{\epsilon=0}$
has vanishing scalar curvature and $S_{\epsilon=0}$ is totally geodesic).
Moreover 
$R^{S_{\epsilon}} = 1/(2M^2) + O(\epsilon)$. Thus,
\begin{eqnarray*}
|S'_{\epsilon} |= |S_{\epsilon}| + \epsilon^2 \left\{ \int_{U_{\epsilon}}
\left [ \frac{3|x|\xi}{M\sqrt{e}}+ \left(\frac{|\nabla_{S_\epsilon}\xi|^2}{2}+\frac{\xi^2}{8M^2} \right) \right ] \bm{\eta_{S_{\epsilon}}} \right\}+ O(\epsilon^3).
\end{eqnarray*}
It follows that, for small enough $\epsilon$, the area of $S'_{\epsilon}$ is
larger than $S_{\epsilon}$ contrarily to our assumption. This proves 
Theorem \ref{theorem} and, therefore, the existence of 
counterexamples to the  version (\ref{penroseBK}) of the 
Penrose inequality.

A final remark is in order. As already mentioned at the beginning,
the existence of this counterexample does not
invalidate the approach suggested by Bray and Khuri based on the generalized Jang equation to
study the general Penrose inequality. It means, however, that the emphasis should not be put on 
generalized apparent horizons. It may be that the approach can serve to prove the standard
version (\ref{penrose1}) as recently discussed in \cite{BrayKhuri2009-2}. Alternatively, 
let us note that, since the slice $\Sigma_{\epsilon}$ lies in the Kruskal
spacetime, it is immediate that the generalized Jang equation admits solutions on $\Sigma_{\epsilon}$
which blow up a non-empty subset of $\partial \Sigma^{+}_{\epsilon}$ and blown down
on another non-empty subset of this boundary, provided the warping function $\varphi^2$ is chosen to be
$\varphi^2 = 1 - 2M/r |_{\Sigma_{\epsilon}}$.
The induced metric on the graph
is then isometric to the Schwarzschild metric $h = \frac{dr^2}{1- 2M/r} + r^2 d \Omega^2$ restricted
to $r > 2M$. The boundary is therefore a minimal surface (in fact, totally geodesic) despite the fact
that $\partial \Sigma_{\epsilon}$ is not smooth in $\Sigma_{\epsilon}$.
This property turns out to be  general
for any  slice $\Sigma$ in an asymptotically flat spacetime with a hypersurface orthogonal Killing
vector $\vec{\eta}$ which is timelike at infinity \cite{Chrusciel1999}.
More precisely, assuming $(\Sigma,g,K)$ to be analytic and defining $\Sigma^{+}$ to be
the largest connected subset of $\Sigma$ containing the asymptotic end such that $\vec{\eta}$ is timelike, the
so-called quotient metric $h$ can be defined on $\Sigma^{+}$. In general, 
$\partial \Sigma^{+}$ is not smooth. However, there exists a differentiable structure on
$\overline{\Sigma^{+}}$ such that $\partial \Sigma^{+}$ is smooth and either lies at infinity
with respect to $h$, or else, this metric 
extends smoothly to the boundary, which 
becomes a totally geodesic submanifold \cite{Chrusciel1999}.
This fact seems to suggest that the PDE method
of Bray and Khuri might 
be suitable even for approaching the second inequality 
in (\ref{penrose2}). At present, however, this remains rather speculative.


\appendix
\section{Proof of Proposition \ref{proposition}}

Firstly, let us consider surfaces in $\Sigma_{\epsilon}$ defined by 
$\left\{ \ycoor=y(x,\epsilon),x,\phi \right\}$ such that the embedding function has the form
$\y = \epsilon \Y$, where $\Y \in U^{2,\alpha}$. An explicit computation of the mean
curvature $p$ on such surfaces gives
$p = \epsilon \Pcal (\Y(x),\dot{\Y}(x),\ddot{\Y}(x),x, \epsilon)$,
where dot denotes derivative with respect to $x$ and where
$\Pcal : \mathbb{R}^3 \times[-1,1] \times I \rightarrow \mathbb{R}$ is a smooth (in fact, analytic) 
function. Similarly $q = \epsilon \Qcal (\Y(x),\dot{\Y}(x),x, \epsilon)$, where 
$\Qcal : \mathbb{R}^2 \times[-1,1] \times I \rightarrow \mathbb{R}$ is an analytic function. Moreover,
the function $\Qcal$ has the symmetry $\Qcal\left(x_1,x_2,x_3,x_4\right) = -
\Qcal\left(x_1,-x_2,-x_3,x_4\right)$, which reflects the fact that the extrinsic curvature
of $\Sigma_{\epsilon}$ changes sign under a transformation $x \rightarrow -x$. Let us write
$\P(Y,\epsilon)(x) \equiv \Pcal(\Y(x),\dot{\Y}(x),\ddot{\Y}(x),x,\epsilon)$ and
similarly $\Q(Y,\epsilon)(x) \equiv \Qcal (\Y(x),\dot{\Y}(x),x,\epsilon)$.

Now, instead of $f$, let us consider 
the functional $\F : U^{2,\alpha} \times I \rightarrow U^{0,\alpha}$ 
defined by $\F (\Y,\epsilon)= \P (\Y,\epsilon) - 
|\Q (\Y,\epsilon)|$. This functional has the property that, for
$\epsilon>0$, the solutions of $\F(\Y,\epsilon)=0$ correspond exactly
to the solutions of $\f(\y,\epsilon)=0$
via the relation $\y = \epsilon \Y$.
Moreover, the functional $\F$ is well-defined for all $\epsilon \in I$, in
particular at $\epsilon=0$.  
Therefore, by proving that $\F=0$ admits solutions 
in a neighbourhood of $\epsilon=0$, we will conclude that $\f=0$
admits solutions for $\epsilon > 0$ and the solutions will in fact belong to a 
neighbourhood of $\y=0$ since $\y = \epsilon \Y$. 

In order to show that $\F$ admits solutions we will use the implicit function
theorem. A direct calculation yields $\F (\Y,\epsilon=0)(x) = c
\left ( L(\Y)(x) -  3|x| \right)$
where $c$ is the constant $1/(m\sqrt{e})$ and
$L(\Y) \equiv -(1-x^2)\ddot{\Y} +2x\dot{\Y}+\Y$. This operator is an isomorphism
between $U^{2,\alpha}$ and $U^{0,\alpha}$. Let $\Ysol \in U^{2,\alpha}$ be the unique
solution of the equation $L(\Y)=3|x|$. For later use, we note that $\Q(\Ysol,\epsilon=0) = 
-3 c x$. This vanishes {\it only} at $x=0$. This is the key property that allows
us to prove that $\F$ is $C^1(U^{2,\alpha}\times I)$.

The $C^1(U^{2,\alpha}\times I)$ property of the functional $\P (\Y,\epsilon)$ is standard.
More subtle is to show that $|\Q|$ is $C^1(U^{2,\alpha}\times I)$ in a
suitable neighbourhood of $(\Ysol,\epsilon=0)$. Let $r_0 >0$ and 
define ${\cal V}_{r_0} = \{ (\Y,\epsilon)\in U^{2,\alpha}\times I:
\| (\Y-\Ysol,\epsilon) \|_{U^{2,\alpha}\times I} \leq r_0 \}$. 
First of all we need
to show that $|\Q|$ is (Fr\'echet-)differentiable on ${\cal V}_{r_0}$, i.e.
that for all $(\Y,\epsilon) \in {\cal V}_{r_0}$ 
there exists a continuous linear mapping $D_{\Y,\epsilon}|\Q|:U^{2,\alpha}\times
I \rightarrow U^{0,\alpha}$ such
that, for all $(H,\delta) \in U^{2,\alpha} \times I$,
$|\Q(\Y+H,\epsilon+\delta)|-|\Q(\Y,\epsilon)| =D_{\Y,\epsilon}|\Q| (H,\delta)+R_{\Y,\epsilon}(H,\delta)$
where $\| R_{\Y,\epsilon}(H,\delta)\|_{U^{0,\alpha}}=o(\|(H,\delta)\|_{U^{2,\alpha}\times I})$. 
The key observation is that, by choosing $r_0$ small enough, we have 
\begin{eqnarray}
|\Q(\Y,\epsilon)(x)|=-\sigma(x)\Q(\Y,\epsilon)(x) \label{key}
\end{eqnarray}
where $\sigma(x)$ is the {\it sign}
function, (i.e. $\sigma(x)=+1$ for $x \geq 0$ and $\sigma(x)=-1$ for $x<0$).
For $x$ away from a neightbourhood of $0$, this is a consequence of the fact that
$\Q(\Ysol,\epsilon=0) = -3 c x$, which is negative for $x>0$ and positive for $x<0$.
Taking $r_0$ small enough, and using that $\Qcal$ is a smooth function of their arguments, the same inequalities hold
for any $(\Y,\epsilon) \in {\cal V}_{r_0}$. Moreover, the function $\Q (\Y,\epsilon)(x)$
is odd in $x$, so it passes through zero at $x=0$. Hence, in a small enough neighbourhood of $x=0$,
the relation (\ref{key})
holds provided we can prove that $\Q(\Y,\epsilon)$ is strictly decreasing at $x=0$.
But this follows inmediately from the fact that $\frac{d \Q (\Ysol,\epsilon=0)}{dx} |_{x=0} = -3 c$ and
$\Qcal$ is a smooth function of its arguments.

From its definition, it follows that $\Q (\Y,\epsilon)(x)$ is $C^{1,\alpha}$ and 
that the functional $\Q_{\Y,\epsilon}$ has 
Fr\'echet derivative 
$D_{\Y,\epsilon}\Q(H,\delta)(x)=A_{\Y,\epsilon}(x)H(x)+B_{\Y,\epsilon}(x)\dot{H}(x)+
C_{\Y,\epsilon}(x)\delta$, where $A_{\Y,\epsilon}(x) \equiv \partial_{1} \Qcal |_{(\Y(x),\dot{\Y}(x),x,\epsilon)}$,
$B_{\Y,\epsilon}(x)  \equiv \partial_{2} \Qcal |_{(\Y(x),\dot{\Y}(x),x,\epsilon)}$ and
$C_{\Y,\epsilon}(x)  \equiv  \partial_{4} \Qcal |_{(\Y(x),\dot{\Y}(x),x,\epsilon)}$. We note that these three
functions are $C^{1,\alpha}$ and that $A_{\Y,\epsilon}$, $C_{\Y,\epsilon}$ are odd, while
$B_{\Y,\epsilon}$ is even (as a consequence of the symmetries of $\Qcal$). Defining the linear map
$D_{\Y,\epsilon}|\Q| (H,\delta) \equiv -\sigma( A_{\Y,\epsilon}H+B_{\Y,\epsilon}\dot{H}+C_{\Y,\epsilon}\delta )$,
it follows from (\ref{key}) that
$|\Q(\Y+H,\epsilon+\delta)|-|\Q(\Y,\epsilon)|= D_{\Y,\epsilon} |\Q| (H,\delta) + R_{\Y,\epsilon}
(H,\delta)$ with $\| R(H,\delta)\|_{U^{0,\alpha}}=o(\|(H,\delta)\|_{U^{2,\alpha}\times I})$. 
In order to conclude that $D_{\Y,\epsilon} |\Q|$ is the derivative of $|\Q(\Y,\epsilon)|$,
 we only need to check that, it is (i) well-defined (i.e. that its image belongs to $U^{0,\alpha}$) and (ii)
that it is continuous, i.e. that $\| D_{\Y,\epsilon} |\Q| (H,\delta) 
\|_{U^{0,\alpha}}
< C \|(H,\delta) \|_{U^{2,\alpha}\times I}$
for some constant $C$. To show (i), the most difficult term is
$-\sigma B_{\Y,\epsilon} \dot{H}$, because $B_{\Y,\epsilon}(x)$ is even and need not vanish at $x=0$.
However $\dot{H}$ is an odd function, and hence $-\sigma B_{\Y,\epsilon} \dot{H}$ is continuous.
To show it is also H\"older continuous, we only need to consider points $x_1 = -a$ and $x_2 = b$
with $0 < a < b$ (if $x_1\cdot x_2\geq 0$, the sign function remains constant, so 
$-\sigma B_{\Y,\epsilon} \dot{H}$ is in fact $C^{1,\alpha}$). Calling $w(x) \equiv
-\sigma(x) B_{Y,\epsilon} (x) \dot{H}(x)$
and using that $w(x)$ is even, we find
\begin{eqnarray}
\hspace{-25mm} |w(x_2) - w(x_1)| = |w(b) - w(-a)| = |(w(b) - w(a) | =
\left |\left . 
\frac{d (B_{Y,\epsilon} \dot{H})}{dx} \right |_{x=\zeta} \right |
|b-a| \leq \nonumber \\
\leq \left | \left . \frac{d (B_{Y,\epsilon} \dot{H})}{dx} \right |_{x=\zeta}
\right | 
|b-a|^{1-\alpha}
|x_2 - x_1|^{\alpha} 
\leq 
\left | \left . 
\frac{d (B_{Y,\epsilon} \dot{H})}{dx} \right |_{x=\zeta} \right |
|x_2 - x_1|^{\alpha}.
\label{estim}
\end{eqnarray}
where $\zeta \in (a,b)$ and we have used that $|b-a|^{\alpha} \leq |b+a|^{\alpha} = |x_2 - x_1|^{\alpha}$
and $|b-a| < 1$. This proves that $- \sigma \B \dot{H}$ is H\"older continuous with exponent $\alpha$.

To check (ii), we first notice that
$w(x)$ obsviously satisfies $\sup_{x} |w| < C \| (H,\delta) \|_{U^{2,\alpha}\times I}$ because 
$B_{\Y,\epsilon}(x)$ is $C^{1,\alpha}$. It remains
to bound the H\"older constant 
$[w]_{\alpha} \equiv \sup_{x_1
\neq x_2} \frac{| w(x_2) - w(x_1)|}{|x_2 - x_1|^{\alpha}}$. Combining 
(\ref{estim}) with the fact that $B_{\Y,\epsilon}(x)$ is $C^{1,\alpha}$, the bound
$[w]_{\alpha}  \leq C \| (H,\delta) \|_{U^{2,\alpha}\times I}$
follows at once. This proves (ii) for the term $-\sigma B_{Y,\epsilon} \dot{H}$.
A similar argument applies to $- \sigma A_{\Y,\epsilon} H$ 
and $- \sigma C_{\Y,\epsilon} \delta$ and we conclude that
$D_{\Y,\epsilon} |\Q|$ is indeed a continuous operator.

In order to apply the implicit function theorem, it is furthermore necessary 
that $|\Q|\in C^1(U^{2,\alpha}\times I)$ (i.e. that 
$D_{Y,\epsilon} |\Q|$ depends continuously on $(\Y,\epsilon)$). 
This means that given any convergent sequence $(\Y_n,\epsilon_{n})\in 
{\cal V}_{r_0}$, the corresponding operators $D_{\Y_n,\epsilon_n} |\Q|$ also
converge. Denoting by $(\Y,\epsilon) \in {\cal V}_{r_0}$ 
the limit of the sequence, we need to 
prove that $\| D_{\Y_n,\epsilon_n} |\Q| - D_{\Y,\epsilon} |\Q|
\|_{\pounds (U^{2,\alpha} \times I, U^{0,\alpha})} \rightarrow 0$. It suffices to
find a constant $K$ (which may depend on $(\Y,\epsilon)$), such that
\begin{eqnarray}
\hspace{-15mm} \| ( D_{\Y_n,\epsilon_n} |\Q| - 
D_{\Y,\epsilon} |\Q| ) (H,\delta) \|_{U^{0,\alpha}}< 
K \|(H,\delta)\|_{U^{2,\alpha}\times I} \|( \Y_n-\Y,\epsilon_{n}-\epsilon)\|_{U^{2,\alpha} \times I}
\label{convergence}
\end{eqnarray}
for all $(H,\delta)\in U^{2,\alpha} \times I$. Again, the most difficult case
involves $\sigma (B_{Y,\epsilon} - B_{Y_n,\epsilon_n}) \dot{H}$, so we concentrate on this term. Using the mean value theorem on the function
${\cal B} \equiv \partial_2 \Qcal$ (recall that $B_{Y,\epsilon}(x) = {\cal B} |_{(Y(x),\dot{Y}(x),x,
\epsilon)}$) gives
\begin{eqnarray}
\hspace{-15mm} \sup_x | \sigma 
({B}_{\Y,\epsilon} - {B}_{\Y_n,\epsilon_n}) \dot{H} | \leq  
2 \sup_{\mathbb{K}}|\nabla {\cal B}|
\sup_x | \dot{H} | \| (Y_n  - Y,\epsilon_{n}-\epsilon) \|_{U^{2,\alpha}\times I}, 
\label{ineq1}
\end{eqnarray}
where 
$\nabla {\cal B}$ is the gradient of ${\cal B}$ and 
$\mathbb{K} \subset \mathbb{R}^4$
is a compact domain depending only on $r_0$ and $\Ysol$ defined so that,
for all $(\Y,\epsilon) \in {\cal V}_{r_0}$, the quadruple
$(\Y(x), \dot{\Y}(x),x,\epsilon) \in \mathbb{K}$, for all $x \in [-1,1]$.
Inequality (\ref{ineq1}) is already of the form (\ref{convergence}) (recall that ${\cal B}$ is smooth). 
It only remains to bound the H\"older constant
of $z \equiv \sigma (B_{\Y,\epsilon} - B_{\Y_n,\epsilon_n}) \dot{H}$ in a similar way.
As before,
this is done
by distinguishing two cases, namely when $x_1 \cdot x_2 \geq 0$ and when 
$x_1 \cdot x_2 <0$. Obtaining an inequality of the form 
$\sup_{x_1 \neq x_2, x_1 \cdot x_2 \geq 0} \frac{| z(x_2) - z(x_1)|}{|x_2 - x_1|^{\alpha}} \leq K_1 \|(H,\delta)\|_{U^{2,\alpha}\times I} \|(\Y_n-\Y,\epsilon_{n}-\epsilon)\|_{U^{2,\alpha} \times I}$ is standard, because $\sigma(x)$ is a constant function.
When $x_1 \cdot x_2 <0$, we exploit the parity of the functions 
as in (\ref{estim}) to get
$| z(x_2) - z(x_1) | \leq 
\left | \left .  \frac{d ((B_{\Y_n,\epsilon_n} - B_{\Y,\epsilon}) \dot{H})}{dx}
\right |_{x=\zeta}  \right |
|x_2 - x_1|^{\alpha},$
where $\zeta \in (a,b)$ and we are asumming $x_1 = -a, x_2 = b, 0 < a < b$
without loss of generality. Bounding the right hand side
in terms of $K_2
\|(H,\delta)\|_{U^{2,\alpha}\times I} \|(\Y_n-\Y,\epsilon_{n}-\epsilon)\|_{U^{2,\alpha} \times I} |x_2 - x_1|^{\alpha}$
 is again standard, since the sign function $\sigma(x)$ has already
disappeared. This, combined with (\ref{ineq1}) gives (\ref{convergence}) and hence
continuity of the derivative of $D_{\Y,\epsilon} |\Q|$ with respect to
$(\Y,\epsilon) \in {\cal V}_{r_0}$.

The final requirement to apply the implicit function theorem to $\F = \P  - |\Q|$  is to 
check that $D_{\Y} \F |_{(\Ysol,\epsilon=0)}$ is invertible. A simple computation
gives $D_\Y \F |_{(\Ysol,\epsilon=0)} (H) = c L (H)$, where
$L$ is the elliptic operator defined above, which is 
an isomorphism between $U^{2,\alpha}$ and $U^{0,\alpha}$. Thus, the implicit function theorem
can be used to conclude that there exists an open
neighbourhood $\tilde {I} \subset I$ of $\epsilon=0$ and a $C^1$ map 
$\tilde{Y}: \tilde{I} \rightarrow U^{2,\alpha}$
such that $\tilde{Y} (\epsilon=0) = \Ysol$ and $\y = \epsilon 
\tilde{Y}(\epsilon)$ defines a $C^{2,\alpha}$ generalized
apparent horizon embedded in $\Sigma_{\epsilon}$.  This proves
Proposition \ref{proposition}.


\section*{Acknowledgments}

We are very grateful to M. S\'anchez for his interest and inestimable help
and to H.L. Bray and M. Khuri for useful comments.
Financial support under the projects
FIS2009-07238 (Spanish MEC), GR-234 (Junta de Castilla y Le\'on) 
and P09-FQM-4496 (Junta de Andaluc\'{\i}a and FEDER funds) are acknowledged.
AC acknowledges the Ph.D. grant AP2005-1195 (MEC).

\end{document}